\documentclass[prd,showpacs,showkeys,twocolumn,floatfix]{revtex4} 
\usepackage[]{epsfig,dcolumn}
\usepackage{graphicx}
\usepackage{bm}

\def\x{{\bf x}}
\def\y{{\bf y}}

\def\k{{\bf k}}
\def\q{{\bf q}}
\def\p{{\bf p}}

\def\A{{\bf A}}
\def\B{{\bf B}}

\def\lsim{\mathrel{\rlap{\lower4pt\hbox{\hskip1pt$\sim$}}
    \raise1pt\hbox{$<$}}}
\def\gsim{\mathrel{\rlap{\lower4pt\hbox{\hskip1pt$\sim$}}
    \raise1pt\hbox{$>$}}}
\begin{document}

%\preprint{IFT/12/02}

\title{ Confinement and gluon propagator in Coulomb gauge QCD} 

\author{ Adam P. Szczepaniak }
\affiliation{ Physics Department and Nuclear Theory Center \\
Indiana University, Bloomington, Indiana 47405 }

\date{\today}

\begin{abstract}
We consider the effects of the Faddeev-Popov determinant in the Coulomb
gauge on the confinement properties of the QCD vacuum. We show that the 
the determinant is needed to regularize the otherwise divergent 
 functional integrals near the Gribov horizon but still enables large
 field configurations to generate IR enhanced running coupling. The
 physical  gluon propagator is found to be strongly suppressed in the
 IR consistent with  expectations  from lattice gauge calculations.

\end{abstract}


\maketitle
\section{Introduction} 
 Quantitative understanding of confinement and more generally  
 of the dynamics of gluons at low energies remains as  the 
 major challenge in QCD. In the past few years  lattice simulations 
 and phenomenological studies  
 have provided new insights into the nature of the low energy behavior of the 
 gluon propagator and role of gluons in forming the hadronic
 spectrum~\cite{Alkofer:2000wg, Michael:2003ai}. 
 Since gluons can only participate in strong interactions, 
 spectroscopy of hadrons with excited gluonic modes is of crucial 
 importance for investigations of confinement.  It has recently been 
 shown that hybrid mesons with excited quark and gluon modes should 
 have properties similar to that of  ordinary hadronic resonances and thus 
 gluonic excitations may appear in the  
 meson spectrum~\cite{Juge:1999ie, Cohen:1998jb,
 Isgur:1999kx, Szczepaniak:2001qz, Afanasev:1999rb}. 
 Searches for exotic mesons have produced a few tantalizing
 candidates~\cite{rhopi, etaprim, e3, e4, Dzierba:2003fw, Szczepaniak:2003vg}
 and new experiments planed for JLab and  GSI in light and charm meson 
 spectroscopy, respectively, are expected to produce a map of gluonic 
 excitations. 
In this paper we address gluon propagation in the QCD vacuum. 
   This investigation was prompted by recent lattice results 
  indicating that in both covariant and Coulomb gauges the low momentum 
 gluons do not propagate~\cite{Bonnet:2000kw, Bowman:2002fe, Cucchieri:2000gu, Cucchieri:2000kw, Dan1}. 
   This is precisely what one expects for 
  physical degrees of freedom {\it e.g.} two transverse gluons in 
  a physical 
 gauge~\cite{Cucchieri:1996ja, Zwanziger:gz, Szczepaniak:1995cw, ss7}. 
 In the four-dimensional  Euclidean formulation of 
 covariant gauge QCD, however, the lack of IR enhancement in the gluon 
 propagator contradicts the naive expectation that the color confining
   force could   be simply related to the gluon propagator. 
  A popular, phenomenological  approach to gluon (and quark) low energy 
 dynamics is based on a truncation of the self-consistent set of 
 Dyson-Schwinger equations. In many such approaches the gluon
  propagator plays a 
  central role in providing the effective interaction between quarks,
  for example, it is used to generate dynamical chiral symmetry
  breaking~\cite{Roberts:dr}. A soft gluon propagator implies that 
 confinement has to be described by other means. For example it has
  been argued that in covariant gauges the Kugo-Ojima confinement
  criterion for  absence of colored non-singlets in the physical
  spectrum can be  satisfied with a soft gluon propagator and an
  enhanced ghost 
 propagator~\cite{Alkofer:2000wg, vonSmekal:1997is, Watson:2001yv, Zwanziger:2002sh, Zwanziger:2001kw}. 
 We will show that this also seems to be the case in Coulomb gauge
 formulation. 

In a covariant formulation one sacrifices positivity constraints and
 the Fock space representation and introduces additional
(ghost) non-physical degrees of freedom. Alternatively, by relaxing the
requirement of manifest Lorentz covariance it is possible to eliminate 
 all non-physical components and study confinement and other low 
 energy phenomena within  the framework of quantum mechanical wave
 functions. Such an approach has obvious, important 
 implications for quark model based phenomenology. Furthermore, 
 at finite density it allows for well established, diagrammatic, 
 many-body  techniques to be used. 
  A many-body approach has  proven to be successful in treating 
 a variety of low energy phenomena in  QCD. For example the random
 phase  approximation, which is typically  ({\it e.g.} for electron
 gas)  relevant at high-densities, in QCD is also applicable at 
 low-densities and may result in a self-consistent realization of
 confinement~\cite{ss7, Swift:za, Szczepaniak:2000uf}. Due to the 
 long-range nature of the confining interaction, 
 at low-densities quasiparticle excitations  have infinite energy 
 which eliminates colored states from the 
 physical spectrum. Due to the presence of bare quark-antiquark pairs near
 the Fermi-Dirac surface, the  quasiparticle vacuum breaks chiral
  symmetry  and leads to a non-vanishing scalar quark
  density. The collective excitations of this 
 quark-antiquark plasma correspond to the Goldstone bosons~\cite{Finger:gm, Adler:1984ri, LeYaouanc:1984dr, Bicudo:sh,
 Szczepaniak:1996gb, Szczepaniak:2002ir}. 

 The picture described above relies 
 on the existence of a long range, confined quark-quark interaction. Such
 an interaction is expected to arise from the Coulomb operator
 which in the Coulomb gauge Hamiltonian describes direct
 interactions between (color) charge densities. Unlike QED, where this
 interaction is simply determined by the distance between  sources,
 in QCD it is a complicated function of the transverse gluon field and
 can not be thought of as a simple potential {\it i.e.} of the 
 Cornell type~\cite{Szczepaniak:1996tk}. The conjecture that the
 Coulomb operator is  related to the confining interaction is based on
 the observation that it is positive definite and vanishes at the
 Gribov horizon. The Grivbov horizon defines the boundary of the gluon 
 field domain. A number of approximations have been developed to calculate the
 expectation value of the Coulomb operator and to verify this
 conjecture~\cite{ss7, Swift:za, Cutkosky:1983dh, Schutte:sd, vanBaal:1997gu}.
 In the process it 
 has been realized that the Gribov region still contains physically 
 equivalent field  configurations.  To what extent the necessary 
 identification of the wave functional at 
 these gauge-equivalent points  modifies the expectation value of the 
 Coulomb operator  remains an open  issue~\cite{vanBaal:1997gu}. 

  The standard approach to a many body system is to introduce a
  physicaly motivated ansatz for the wave functional 
 and to define approximations for the evaluation of expectation  values. 
  We have followed this
  approach in Ref.~\cite{ss7} where we generated  the confining
  interaction  but a specific assumption on the  behavior of the gluon 
 propagator at low energies had to be imposed. 
  In particular, the gluon dispersion relation which follows
 from minimizing the vacuum expectation value was solved self-consistently
 together with  the confining interaction but a particular boundary
 condition was chosen for the solutions. In absence of the
  Faddeev-Popov determinant this was necessary in order to obtain a
  nontrivial solution to the coupled integral equations. 
  In this paper we will show how the   Faddeev-Popov determinant can
  be  included, how it fixes the low energy gluon dispersion 
 relation and brings it in a qualitative agreement with 
 lattice results. 

 The paper is organized as follows: in Sec.II we  give a brief
 description of QCD in the Coulomb gauge.  As discussed above the main 
 novel feature of this approach is the inclusion of the Faddeev-Popov 
 determinant. The details of the many-body
 formulation are given in Sec.III.  Renormalization is discussed in
 Sec.IV and is followed by numerical results in Sec.V. 

\section{QCD in the Coulomb gauge}

The QCD Coulomb gauge Hamiltonian, 
 defined by $\bm{\nabla}\cdot{A}^a({\bf x})$  is  given by~\cite{Christ:ku},
\begin{equation}
H = H\left[ {\bf A}^a(\x), \bm{\Pi}^a(\x)\right] = 
 H_0 + H_{qg} + H_{g^3} + H_{g^4} + H_C,
\end{equation}
with,  $\bm{\Pi}^a(\x)$ being the canonical momentum satisfying, 
\begin{equation}
[\Pi^a_i(\x),A_j^b(\y)] = -i \delta_{ab}\delta^{ij}_T(\bm{\nabla})\delta^3(\x-\y),
\end{equation}
 $\delta^{ij}_T(\bm{\nabla}) = \delta^{ij} -
 \nabla^i\nabla^j/\bm{\nabla}^2$ and in the Shr{\"o}dinger 
 representation given by $ \bm{\Pi}^a(\x) = -i\delta/\delta {\bf A}^a(\x)$.
 The five terms represent the kinetic energy, the quark-transverse
gluon coupling, the magnetic $3-$ and $4-$ gluon couplings and the 
 instantaneous Coulomb energy,  respectively. In this paper we 
 focus on the gluon sector and thus will ignore quark degrees of 
 freedom. The gluon kinetic term is given by,  

\begin{equation}
H_0 = {1\over 2} \int d^3\x \left[  {\cal J}^{-1}  \bm{\Pi}^a(\x) {\cal J}
{\bm \Pi}^a(\x) + (\bm{\nabla}\cdot \A^a(\x))^2
 \right],
\end{equation}
with 
\begin{equation}
{\cal J} \equiv Det\left(1 - \lambda\right) = e^{Tr\log(1 - \lambda)}, 
\label{J}
\end{equation}
being the Faddeev-Popov (FP) determinant. The matrix $\lambda$, in the
FP operator $(1-\lambda)^{-1}$, is given by,  
\begin{equation}
\lambda_{\x,a;\y,b} =  \left(-
{1\over {\bm{\nabla}^2}} \right)_{\x,\y} g f_{acb}
\A^c(\y)\bm{\nabla}_\y, 
\end{equation}
and in Eq.~(\ref{J}) the trace is over the spatial: $\x,\y$  and
color: $a,b,c$ indices. The FP determinant is  the Jacobian of 
 coordinate transformation from the canonical coordinates of the Weyl gauge, 
$A^{\mu,a}(\x) = (0,{\bf V}^a(\x))$ with kinetic energy given
by, $1/2 \int d^3\x  (-i \delta/ \delta {\bf V}^a(\x))^2$ , 
 to the Coulomb gauge fields, $\A^a(\x)$ defined through a gauge map, 
\begin{equation}
{\bf V}^a(\x) \to \left(\A^a(\x),\vec{\phi}(\x)\right) = 
 u(\vec{\phi}) \A^a u(\vec{\phi}) + u^{-1}(\vec{\phi})
\bm{\nabla} u(\vec{\phi}) \label{map}. 
\end{equation}
The dependence of the Hamiltonian and wave functionals 
 on the $N_c^2 -1$ Euler angles $\vec{\phi}(\x)$ can be eliminated using the
 Gauss's law constraint and results in the Coulomb energy 
term~\cite{Christ:ku}, 
\begin{equation}
H_C = {1\over 2} \int d^3\x d^3\y {\cal J}^{-1} \rho^a(\x) {\cal J}
K[\A]_{\x,a;\y,b}
\rho^b(\y), 
\end{equation}
with $\rho^a(\x)$ being the color charge density, in the absence of
quarks is given by,  
\begin{equation}
\rho^a(\x) = f_{abc}\bm{\Pi}^b(\x)\A^c(\x), 
\end{equation}
and the Coulomb kernel $K$ given by, 
\begin{equation}
K[\A]_{\x,a;\y,b} \equiv g^2  \left[(1 - \lambda)^{-2} \left(-{1\over
 {\bm{\nabla}^2}}\right)\right]_{\x,a;\y,b}. 
\end{equation}
More details of te derivation of the Coulomb gauge can be found in
Ref.~\cite{Christ:ku,ss7}.

Functional integrals in the Coulomb gauge are performed over the 
 measure, $\Pi_{\x,a,i} dA^{i,a}(\x) {\cal J}$. The Faddeev-Popov
 determinant results from the nonlinear field transformation given in
  Eq.~(\ref{map}) and reflects the complicated topology of the field
 space domain.  Furthermore it is well known that the gauge condition, 
  $\bm{\nabla}\cdot {\bf A}^a = 0$ is not a complete gauge
 fixing and thus the mapping $V \to A,\phi$ is not unique. The
 unique solution on a gauge orbit can be defined as   
 the absolute minimum of the functional, 
 $I[A,g]=\int d\x (\A^a(\x))^2_g $ minimized
 over $g$~\cite{vanBaal:1997gu}. At the minimum of $I[A]$, 
 $\bm{\nabla}\cdot\A^a=0$ and the FP operator 
 is positive. The space of the absolute minima defines the
 Fundamental Modular Region, $\Lambda$ as shown in Fig.~1. 
 The  boundary of $\Lambda$ is a set of gauge fields which lead to 
 degenerate absolute minima. The fundamental region resides inside 
 the so called Gribov region, $\Omega$, corresponding to all minima 
 of $I[A]$ and thus also satisfying the transversality condition. The
 boundary $\partial\Omega$ defines a set of configurations for which the
 gauge mapping is singular and  ${\cal J}[A]=0$. 
 In what follows we will primarily study the role of 
 configurations near singular boundary points on $\partial\Omega$. 
 Since there exist configurations for which $\partial \Lambda$ and $\partial
 \Omega$ overlap, the point $A^a(\x) = 0$ lies in $\Lambda$  and both
 $\Lambda$ and $\Omega$ are convex, some fluctuations around the null 
 field reach the coordinate singularity, $\delta\Omega$ without leaving
  $\Lambda$. Furthermore it has been pointed
 out~\cite{Zwanziger:2003cf} that it is the common boundary points 
 which dominate functional 
 integrals. Thus even if field configurations outside of FMR
   are included these may not lead to substantial  errors.

 \begin{figure}
 \includegraphics[width=2.5in]{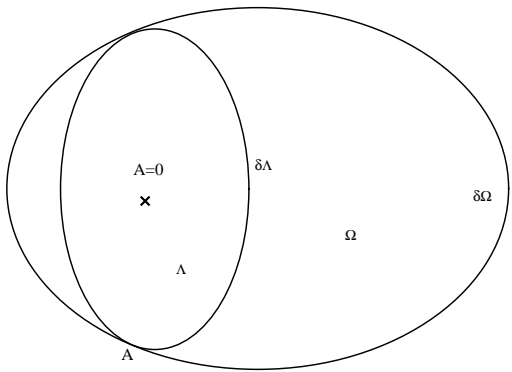}
 \caption{\label{fig1} A schematic representation of the field domain in
   the Coulomb gauge. The point A is on the common boundary of the
 fundamental modular region ($\delta \Lambda$) and the Gribov region
 ($\delta \Omega$) and corresponds to a coordinate singularity, ${\cal
 J}[A] = 0$ }
 \end{figure}

 In the limit ${\cal J}=1$, the kinetic term describes a set of coupled 
  harmonic oscillators and its (unnormalized) 
 ground state $|0\rangle$ is given by 
 \begin{equation}
 \langle \A| \omega_0 \rangle = \exp\left( - {1\over 2}\int 
 {{d^3\k}\over {(2\pi)^3}}
 \omega_0(k) \A^a(\k) \A^a(-\k) \right) \label{0}
 \end{equation}
 with $\omega_0(k) = k = |\k|$ being the free gluon energy and, 
 \begin{equation}
 \A^a(\k) \equiv \int d\x e^{-i\k\cdot \x} \A^a(\x), 
 \end{equation} 
  represents the normal modes. The FP determinant leads to a 
   suppression of the ground state wave functional near the Gribov
   horizon. This can be illustrated using an analogy between the Weyl
   and the Coulomb gauge kinetic terms and a harmonic oscillator 
 in cartesian and spherical coordinates, respectively. In 
   $N$-dimensions, the $S$-wave harmonic oscillator 
  radial wave function satisfies, 
 \begin{equation}
 {1\over 2}\sum_{i=1}^N  \left[ - {{\partial^2}\over {\partial x^2_i}} 
  + \omega^2 x^2_i \right] R(r) = {1\over 2} \left[ {\cal J}  
  {\partial \over
  {\partial r}} {\cal J}  {\partial \over  {\partial r}} + 
  \omega^2 r^2  \right] R(r). 
 \end{equation}
 Here the jacobian is given by 
 ${\cal J} = r^{N-1} \sim  \exp(-N \log r)$ and it vanishes at the
 boundary $r\to 0$ of the domain of $r$. The ground state 
  wave function, $R(r)$  is finite at that boundary, 
  $R(r) = exp(-r^2 \omega^2/2)$ but the radial wave function defined by 
  $u(r) = {\cal J}^{1/2} R(r)$, vanishes as $r \to 0$. 
 The Hamiltonian can be redefined to absorb the jacobian,  
 \begin{equation}
 H \to {\bar H} = {\cal J}^{-1/2} H {\cal J}^{1/2} = 
  {1\over 2}\left( p^2_r + \omega^2 r^2\right) + V_C, \label{ho}
 \end{equation}
 were, $p_r = -i\partial/\partial r$ and the additional potential is given 
   by $V_C = {\cal J}^{-2}[p_r, {\cal J}]^2/4 - {\cal J}^{-1}[p_r,[p_r, 
  {\cal J}]]/2$. The Hamiltonian ${\bar H}$ is hermitian with respect
   to a flat measure in the radial direcction, 
  \begin{equation}
{{ \int D\x R(\x) H R (\x)}\over {\int d\Omega}} = \int dr {\cal J} R(r) H R(r) 
  =  \int dr u(r) {\bar H} u(r).
 \end{equation}
 In terms of ${\bar H}$ and $u(r)$ one effectively 
 recovers the simple harmonic motion in one dimension 
 (modulo $V_C$), except for a boundary condition, $u(r) \to 0$ 
 at a point $r = 0$ corresponding to the singularity of the 
 coordinate transformation.

The behavior of the wave function near the Gribov horizon is strongly
 correlated with confinement. In the Coulomb gauge, at the Gribov
 horizon, ${\cal J} = 0$ the Coulomb kernel diverges and this can be
 interpreted as a manifestation of confinement~\cite{ss7}. However, 
 since the radial wave functional vanishes at singular points, $\delta
 \Omega$, of the coordinate transformation,
  functional integrals over color singlet states are expected to be
 finite. In Ref.~\cite{ss7} we have explored this scenario, but we 
 have not accounted for the boundary condition on the ground state wave
 functional. In effect we used a gaussian ansatz for $u$ and not
  for $R$, {\it i.e.} the wave functional was finite at the Gribov
 horizon. To ensure that functional integrals over the FP operator
 converge we had to choose a particular condition on the parameters of
 the ground state wave functional. 
 Summarizing, the FD determinant is a  crucial element of the QCD 
 Coulomb gauge dynamics  as it surpresses the integrands in functional
 integrals near the Gribov horizon.

 As long as gauge fields  are within the Gribov region, $\Omega$,  
  the Coulomb potential is positive and it is possible to 
  use a variational approach. The simplest variational ansatz for which
 diagrammatic expansion is possible is given by the
 functional generalization of the harmonic oscillator ground state
 which is equivalent to the quasiparticle approximation. 
 This approximation can be systematically  improved through a cluster 
 expansion and excited states can also be  studied~\cite{Szczepaniak:2002ir}. 
 In this context one often uses the formalism of second quantization
 which is natural when dealing with gaussian integrals over polynomials
 (Wick theorem). In our case, however, before one gets to this stage
 one has to deal with non-polynomial operators, {\it e.g.} ${\cal
 J}[A]$ or  $K[A]$, and thus it is simpler to procceed with the Schr{\"
 o}dinger representation.

\section{The quasiparticle spectrum}
In the non-interacting case, $H=H_0$ the perturbative vacuum of 
  Eq.~(\ref{0}) minimizes the energy density of the system, {\it i.e.} 
 \begin{eqnarray}
0= {\partial \over {\partial \omega(k)}} & & 
 {{\langle \omega|H_0| \omega \rangle/{\cal V}}\over {\langle \omega| \omega \rangle
}} _{|\omega=\omega_0} \nonumber \\ 
& &  =  {\partial   \over {\partial \omega(k)}} 
 {1\over 4}\int {{d^3\q}\over {(2\pi)^3}}  \left[ \omega(q) + {{{\bf
 q}^2 } \over {\omega(q)} } \right]_{\omega=\omega_0}.
\end{eqnarray}
 Here ${\cal V}$ is the total number of gluon degrees of freedom, 
${\cal V} = \delta_{aa} \delta_T^{ii} \int d^3\x = 
(N_C^2 - 1)\times 2 \times \mbox{ volume}$. To describe the 
 quasiparticle spectrum we will use the same gaussian variational ansatz.
 The {\it vev} of the Hamiltonian becomes, 
\begin{equation}
E(\omega) = \langle \omega |H\omega\rangle/\langle
\omega|\omega\rangle \equiv E_K(\omega) + E_C(\omega), 
\end{equation}
where 
\begin{widetext}
\begin{equation}
E_K = {1\over {2  {\langle \omega|\omega\rangle}}} 
  \int D\A^a 
 \int d\x e^{-\int d\k{\omega(k)\over 2} 
\A^a(\k)\A^a(-\k)}\left[  \bm{\Pi}^a(\x){\cal J}\bm{\Pi}^a(\x) 
+ ( {\bm\B}^a(x))^2\right]e^{-\int d\k{\omega(k)\over 2} 
\A^a(\k)\A^a(-\k)},
\end{equation}
\end{widetext}
 $\B^a(\x) = \bm{\nabla}\times \A^a(\x) + gf_{abc} \A^b(x)\times
\A^c(\x)/2$, and 
\begin{widetext}
\begin{equation}
E_C = {1\over {2 \langle \omega|\omega\rangle}} 
   \int  D\A^a 
 \int d\x d\y e^{-\int d\k
  {{\omega(k)}\over 2}
\A^a(\k)\A^a(-\k)}\left[ \rho^a(\x) {\cal J} 
 K[\A]_{\x,a;\y,b} \rho^b(\y) \right]e^{-\int d\k{\omega(k)\over 2} 
\A^a(\k)\A^a(-\k)}.
\end{equation}
\end{widetext}
The ground state normalization is given by, 
\begin{equation}
\langle \omega|\omega \rangle =  \int 
  D\A^a {\cal J}[A]  e^{-\int d\k \omega(k) \A^a(\k)\A^a(-\k)}
 \equiv  \langle {\cal J}[A] \rangle 
\end{equation}
 We use the $\langle \cdots \rangle$ to represent  
 functional integrals over the  ground state ansatz functional 
 with the {\it flat} measure. 
 After integrating by parts the kinetic and Coulomb kernel contributions can
 be written as ( $[d\k] \equiv d\k/(2\pi)^3$), 
\begin{eqnarray}
& &  E_K = {1\over { 2\langle \omega|\omega\rangle}} 
   \int  D\A^a 
 {\cal J} 
 e^{-\int d\k \omega(k) 
\A^a(\k)\A^a(-\k)} \nonumber \\
& & \times  \left[ 
 \int [d\k] 
 \omega^2(k) \A^a(\k) A^a(-\k)
 + \int d\x ( {\bm\B}^a(x))^2\right],\nonumber \\ \label{Ek}
\end{eqnarray}
 and 
\begin{eqnarray}
& & E_C = {1\over {2\langle \omega|\omega\rangle}}
   \int  D\A^a 
 \int \Pi_{i}^4 [d\k_i] 
 {\cal J}  e^{-\int d\k \omega(k) 
\A^a(\k)\A^a(-\k)} \nonumber \\
& &  \rho^a(\k_1,\k_2)
K[\A]_{\k_i,a,b} \rho^b(\k_3,\k_4), \nonumber \\
\end{eqnarray}
respectively. The charge density is now given by,  
\begin{equation}
\rho^a(\k_i,\k_j) = f_{abc}\omega(k_j) \A^b(\k_i)\cdot \A^c(\k_j), \label{rho}
\end{equation}
and the Coulomb kernel by, 
\begin{equation}
K[\A]_{\k_i,a,b} = \int d\x d\y e^{i(\k_1+\k_2)\cdot\x}
K[\A]_{\x,a;\y,b} e^{i(\k_3+\k_4)\cdot \y}.
\end{equation}
Even though details of the  boundary of the functional integrals 
 are not known,  the partial integration is presumably 
 justified since the integrand vanishes as $\A \to \infty$ and 
 at the boundary of the Gribov region, ${\cal J} \to 0$. Compared to
 the harmonic oscillator example discussed earlier, the partial integration
  combines contributions from $V_C$ and $p_r^2$ in Eq.~(\ref{ho})
  to the {\it vev} of the Hamiltonian and express 
 them as coordinate space integrals over the gaussian wave function. 
 In our case the complication in evaluating functional integrals 
  over $D\A$ is 
  due to the nonlinear dependence of  ${\cal J}[A]$ and $K[A]$ on the
 FP operator, $(1-\lambda)^{-1} = (1-\lambda[\A])^{-1}$.  
 These integrals are performed 
 by expanding such functionals in powers of $A$, performing Gaussian 
 integrals of over polynomials in $A$ and approximating them by 
 products of two-point correlations. This is will be illustrated in
 particular cases below. To simplify the
 notation, the triplet of  indices representing momentum, color and spin 
 will be denoted by greek letters, {\it e.g.} $\alpha = (\k,a,i)$ and
 a doublet containing a momentum and a color index by ${\bar \alpha} =
 (\k,a)$. The summation convention will be used with upper and lower
 indices differing by a replacement $\k \to -\k$,
 {\it e.g.} $A^{\alpha} \equiv A^{ia}(\k)$,
% The dominant contrinution to functional integrals is expected 
% from the boundry region where, $\lambda[A] \to 1$ and  $K[A] \to \infty$. 
% This should orignate from slowly warying configuraion
%space file configurations, since the high momentum momdes are
%respinsible for assymptotic freedom. In evluation of the function
%integrals we will thus rtain the diagrams whcih contain a the maximum
%number of bare interaction, $1/\bm{\nabla}^2$ lines. 
%Now we will be more specific and give the details of evaluation fo the
%two functional integrals.  
\begin{equation}
\sum_\alpha A^\alpha A^\alpha \equiv A^\alpha A_\alpha = 
\sum_{a}\sum_{ij} \int [dk] A^{ia}(\k)\delta_T^{ij}(\k) A^{ia}(-\k). 
\end{equation}
In this notation $\lambda$ from the FP operator can be written as, 
\begin{eqnarray}  
& & \lambda^{\bar\alpha}\mbox{}_{\bar\beta} = 
 \lambda^{(\p,a)}\mbox{}_{(\q,b)} 
  \equiv \lambda^{\bar \alpha}\mbox{}_{\gamma{\bar \beta}}A^\gamma
 \nonumber \\
& & = g f_{acb}
  \int[d\k](2\pi)^3\delta^3(\p-\k-\q) {{\A^c(\k) \cdot i\q} \over {\p^2}}
 \equiv \lambda^{\bar \alpha}\mbox{}_{\gamma{\bar \beta}}A^\gamma,
 \nonumber \\
\end{eqnarray}
with 
\begin{equation}
\lambda^{\bar \alpha}\mbox{}_{\gamma{\bar \beta}} 
 =  \lambda^{(\p,a)}_{\;\;\;(\k,c,k),(\q,b)}
 = (2\pi)^3 \delta^3(\p-\k-\q) 
{{ig[\delta_T(\k)q]^k}\over {\p^2}}
 f_{acb}. 
\end{equation}
%\begin{equation}
%\sum_\gamma {1\over{2\omega_\gamma}} 
% [ \lambda_\gamma \lambda^\gamma ]^\alpha_\beta =
% \lambda^\alpha\mbox{}_{\gamma\sigma}
%\lambda^{\sigma\gamma}\mbox{}_\beta
%\end{equation}
Evaluation of functional integrals is simplified by introducing
 corresponding generating  functionals, 
%\begin{eqnarray}
%& & Z_F(\alpha_1, \cdots, \alpha_n)  \equiv  \langle A^{\alpha_1} \cdots
% A^{\alpha_2} F[A] \rangle = \nonumber \\
%& & =  \int  \Pi_\beta dA^\beta F[A] A^{\alpha_1} \cdots
% A^{\alpha_n} e^{-\sum_\gamma \omega_\gamma A^\gamma A_\gamma }
%\end{eqnarray}
\begin{equation}
\langle A^{\alpha_1} \cdots
 A^{\alpha_n} F[A] \rangle 
 =  \int  \Pi_\beta dA^\beta F[A] A^{\alpha_1} \cdots
 A^{\alpha_n} e^{-\sum_\gamma \omega_\gamma A^\gamma A_\gamma }
\end{equation}
and since, 
\begin{eqnarray}
 \int \Pi_\beta dA^\beta  F[A]   
 & & e^{\sum_\gamma \left[ -\omega_\gamma A^\gamma A_\gamma  + A^\gamma
    J_\gamma \right]} \nonumber \\
& &  =  e^{ \sum_\gamma {{ J^\gamma J_\gamma}\over
 {4\omega_\gamma }}} \langle F\left[A+{J\over {2\omega}}\right] \rangle 
\end{eqnarray}
we obtain, 
\begin{eqnarray}
\langle A^{\alpha_1} \cdots
& &  A^{\alpha_2}F[A] \rangle = \nonumber \\
%Z_F(\alpha_1,\cdots,\alpha_n) = 
& & = {\delta \over {\delta J_{\alpha_1}}} \cdots 
  {\delta  \over {\delta J_{\alpha_n}}}_{J=0} 
  e^{ \sum_\gamma {{ J^\gamma J_\gamma}\over
 {4\omega_\gamma }}} \langle F\left[A+{J\over {2\omega}}\right]
\rangle.  \nonumber \\
\label{Z}
\end{eqnarray}
In this paper we are primarily concerned with the instantaneous part of the
 transverse gluon propagator, 
\begin{equation}
\Pi^\alpha\mbox{}_\beta \equiv 
  \langle A^\alpha A_\beta J[A]  \rangle/\langle
J[A] \rangle   =  {{\delta^\alpha\mbox{}_\beta}\over {2\Omega_\alpha}}.
\end{equation}
 where the  last equality follows from translational invariance and color
neutrality of the vacuum. In the approximation ${\cal J}=1$, one has 
 $\Omega(k) = \omega(k)$ and one obtains the propagator used in
 Ref.~\cite{ss7}. From Eq.~(\ref{Z}) it follows that,  
\begin{widetext}
\begin{equation}
 \Pi^\alpha\mbox{}_\beta = 
{ \delta \over {\delta J_\alpha}}
{\delta \over {\delta J^\beta}}_{J=0} 
 e^{\sum_\gamma {{ J^\gamma J_\gamma}\over
 {4\omega_\gamma  }}}
 \langle {\cal J}\left[A+{J\over {2\omega}}\right]
\rangle / \langle {\cal J}[A] \rangle 
 =  {1\over {2\omega_\alpha}}\left[
  \delta^\alpha\mbox{}_\beta + 2\omega_\alpha 
   {\delta \over {\delta J_{\alpha}}}
{\delta \over {\delta J^{\beta}}}_{J=0} 
  \langle {\cal J}\left[A+{J\over {2\omega}}\right] \rangle 
  /  \langle {\cal J}[A] \rangle  \right], 
\end{equation}
\end{widetext}
 with the first term representing  propagator the absence of the
 FP determinant (${\cal J} = 1$) and the second term given by, 
\begin{widetext}
\begin{eqnarray}
 {\delta \over {\delta J_{\alpha}}} 
{\delta \over {\delta J^{\beta}}}_{J=0} 
  \langle {\cal J}\left[A+{J\over {2\omega}}\right] \rangle 
/ \langle {\cal J} [A] \rangle &  = & 
 -   \langle  \left[ {{\lambda^\alpha}\over {2\omega_\alpha}}
  (1 - \lambda)^{-1} {{\lambda_\beta}\over {2\omega_\beta}}
  (1 - \lambda)^{-1}
\right]^{\bar \gamma}_{\;\;{\bar \gamma}} {\cal J}[A] \rangle /\langle {\cal J}[A]
\rangle  \nonumber \\
& &   +  \langle  \left[ {{\lambda^\alpha}\over {2\omega_\alpha}}
  (1 - \lambda)^{-1} \right]^{\bar \gamma}_{\;\;{\bar \gamma}}
 \left[ {{\lambda_\beta}\over {2\omega_\beta}}
  (1 - \lambda)^{-1} \right]^{\bar \sigma}_{\;\;{\bar \sigma}}
{\cal J}[A] \rangle /\langle {\cal J}[A]
\rangle, 
\end{eqnarray}
\end{widetext}
where $[\lambda^\alpha]^{\bar \gamma}\mbox{}_{\bar \sigma} \equiv 
 \partial \lambda^{{\bar \gamma}\alpha}\mbox{}_{\bar \sigma}/\partial 
 A_\alpha = 
 \lambda^{{\bar \gamma}\alpha}\mbox{}_{\bar \sigma}$.
  This relation can be represented through an infinite set of coupled
 integral Dyson  equations containing all dressed vertices. As argued in 
 Ref.~\cite{ss7, Swift:za}, however,  vertex corrections give a finite 
 and small
 modification and will be ignored. The dominant contributions in 
 both the  IR and the UV regions of the loop momentum integrals 
 over instantaneous propagators come  from diagrams with a maximal
 number of soft Coulomb lines and a maximal number of {\it primitive} self
 energy loops, respectively. The {\it primitive} self energy is shown in
  Fig.~2  and is given by 
\begin{equation}
 I^{0{\bar \alpha}}\mbox{}_{\bar \beta} \equiv 
\sum_\gamma {1\over{2\omega_\gamma}} 
  [ \lambda_\gamma \lambda^\gamma ]^{\bar \alpha}\mbox{}_{\bar \beta} =
 \sum_\gamma {1\over {2\omega_\gamma}} 
\lambda^{\bar \alpha}\mbox{}_{\gamma{\bar \sigma}}
 \lambda^{{\bar \sigma}\gamma}\mbox{}_{\bar \beta}
 =  \delta^{\bar \alpha}\mbox{}_{\bar \beta} I^0_{\bar \alpha},
\end{equation}
 \begin{figure}
 \includegraphics[width=2.5in]{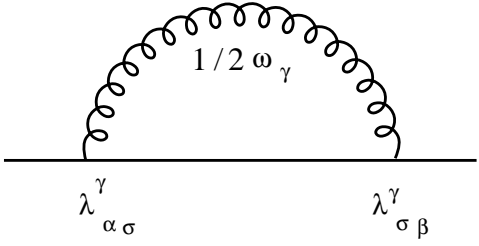}
 \caption{\label{fig2} Primitive self-energy, {\it i.e.} 
 the lowest order correction to the FP
   operator, or the Coulomb line, $I^0$. The solid line represents the
   bare Coulomb potential, $(1/p^2)$.} 
 \end{figure}
and explicitly,  
\begin{equation}
I^0_{{\bar \alpha}} = I^0(q) = g^2 N_C \int [d\k] 
  {{(1 - (\hat{\q}\cdot\hat{\k})^2)}
  \over { 2\omega(|\k|)(\q-\k)^2}} .
\end{equation}
 This self-energy is UV divergent and has to be renormalized. We will
 discuss renormalization in detail in the following section. 
 To proceed we need to introduce the expectation value of the FP
 operator, 
\begin{equation}
d^{\bar \alpha}\mbox{}_{\bar \beta} = \langle [(1 -
 \lambda)^{-1}]^{\bar \alpha}\mbox{}_{\bar \beta}
 {\cal J}[A]\rangle/\langle 
{\cal J}[A]\rangle = \delta^{\bar \alpha}\mbox{}_{\bar \beta} d_{\bar \alpha}.
\end{equation}
 A few lowest order diagram contributions to this {\it vev} 
 are shown in Fig.~3. 
 \begin{figure}
 \includegraphics[width=2.5in]{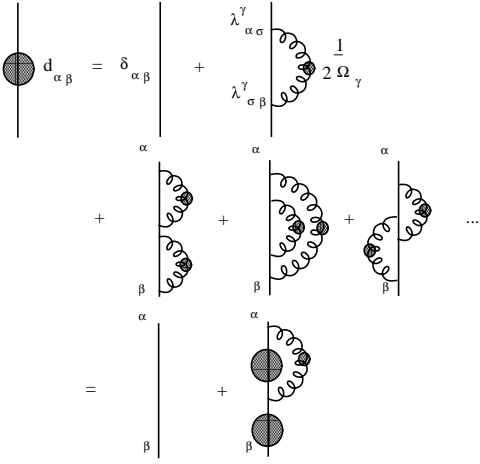}
 \caption{\label{fig3} Rainbow-ladder approximation to the Dyson
   equation for the Faddeev-Popov operator} 
 \end{figure}
At the two-loop order the first and second diagram
 in the second line dominate in the
 IR and UV, respectively and are retained. In higher orders the dominant
 contribution comes from a series of  rainbow-ladder diagrams
 obtained by summing the class of diagrams generated by these
 two  lowest order loop diagrams and results in the following 
 approximation to the Dyson series, 
\begin{eqnarray}
 d^{\bar \alpha}\mbox{}_{\bar \beta }
 & =&  \delta^{\bar \alpha}\mbox{}_{\bar \beta} d_{\bar \alpha} 
  =   \delta^{\bar \alpha}\mbox{}_{\bar \beta }
  + \langle [\lambda (1 - \lambda)]^{\bar \alpha}\mbox{}_{\bar \beta }
  {\cal J} \rangle /\langle {\cal J} \rangle \nonumber \\
& &  = \delta^{\bar \alpha}\mbox{}_{\bar \beta}  + 
\sum_{\gamma} {1\over {2\Omega_\gamma}} 
 [\lambda_\gamma d \lambda^\gamma d ]^{\bar \alpha}\mbox{}_{\bar \beta}
 = \delta^{\bar \alpha}\mbox{}_{\bar \beta} \left[1 + I_{\bar \alpha}
 d_{\bar \alpha} \right], \nonumber \\
\end{eqnarray}
where 
\begin{equation}
\sum_{\gamma}{1\over {2\Omega_\gamma}} 
  \lambda^{\bar \alpha}\mbox{}_{\gamma{\bar \sigma}} d_{\bar \sigma} 
\lambda^{{\bar \sigma}\gamma}\mbox{}_{\bar \beta}
 = \delta^{\bar \alpha}\mbox{}_{\bar \beta} I_{\bar \alpha},
\end{equation}
and explicitly, with  $d_{\bar \alpha} = d(q)$, $I_{\bar \alpha} =
I(q)$, 
is given by 
\begin{eqnarray}
& & d(q) = {g\over {1 - gI(q)}}, \nonumber \\
& &  I(q) = {N_C} \int [d\k] 
  {{(1 - (\hat{\q}\cdot\hat{\k})^2)}
  \over { 2\Omega(|\k|)(\q-\k)^2}}d(|\q-\k|). \nonumber \\ \label{dd}
\end{eqnarray}
Using the same approximation (of ignoring vertex corrections), the
 Dyson equation for the instantaneous propagator becomes, 
\begin{widetext}
 \begin{equation}
{\delta \over {\delta J^{\alpha}}} 
{\delta \over {\delta J_{\beta}}}_{J=0} 
  \langle {\cal J}\left[A+{J\over {2\omega}}\right] \rangle 
/ \langle {\cal J} [A] \rangle  =  
     - \left[  {{\lambda^\alpha}\over {2\omega_\alpha}} 
        d {{ \lambda_\beta}\over {2\omega_\beta}} d \right]^{\bar
        \gamma}_{\;\;{\bar \gamma}}
 + \sum_\rho  \left[{{\lambda^\alpha}\over {2\omega_\alpha}} 
 d \lambda_\rho d \right]^{\bar \gamma}_{\;\;{\bar \gamma}} 
 {1\over {2\Omega_\rho}}
 \left[ \lambda^\rho d {{\lambda_\beta}\over {2\omega_\beta}} d
 \right]^{\bar \sigma}_{\;\;{\bar \sigma}},
\end{equation}
\end{widetext}
and is shown in Fig.~4.  Since neutrality of the vacuum implies 
\begin{equation}
  [\lambda^\alpha d \lambda_\beta d]^{\bar \gamma}\mbox{}_{\bar
 \gamma} = 
\sum_{{\bar \gamma}{\bar \sigma}}
\lambda^{{\bar \gamma}\alpha}\mbox{}_{\bar\sigma}
d_{\bar \sigma} \lambda^{\bar \sigma}\mbox{}_{\beta{\bar \gamma}} 
 d_{\bar \gamma} 
 = 2 \delta^\alpha\mbox{}_\beta  F_\alpha,
\end{equation}
we finally obtain, 
\begin{equation}
\Omega_\alpha = 
\omega_\alpha  + F_\alpha, \label{Omega}
\end{equation}
where $F_\alpha = F(q)$, is explicitly given by
\begin{equation}
F(q) \equiv  {N_C\over 2}  \int [d\k]  {{(1 - (\hat\k\cdot\hat\q)^2)}\over
  {(\q-\k)^2}} d(|\k|) d(|\q-\k|). \label{Omega1}
 \end{equation}
 \begin{figure}
 \includegraphics[width=2.5in]{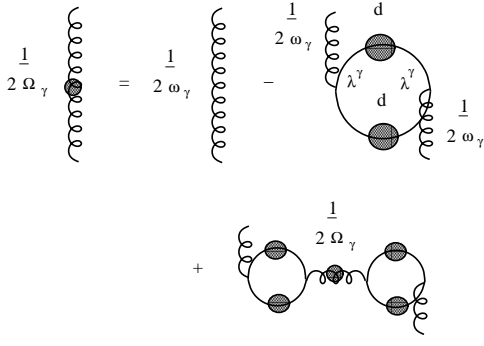}
 \caption{\label{fig4} The rainbow-ladder approximation to the Dyson
   equation for the transverse gluon propagator}
 \end{figure}

We can now return to the calculation of the vacuum expectation of the
 full Hamiltonian. By minimizing with respect to $\omega$ this
 determines  $\omega$, and from Eq.~(\ref{Omega}), the  
 gluon propagator, $1/2\Omega$.   In terms of this propagator, 
 the kinetic vacuum expectation value, $E_K$ is given by 
\begin{eqnarray}
& & E_K =  {1\over 2} \sum_\alpha 
  (\omega_\alpha^2 + p_\alpha^2)\Pi^\alpha\mbox{}_\alpha +
 \nonumber \\
& + &  
  {1\over 2} V_{\sigma\alpha\beta} V^\sigma\mbox{}_{\gamma\delta}   
   {\partial \over {\partial J_\alpha}} 
 {\partial \over {\partial J_\beta}} 
 {\partial \over {\partial J_\gamma}} 
 {\partial \over {\partial J_\delta}} 
  e^{ \sum_\rho {{ J^\rho J_\rho}\over
 {4\omega_\rho }}} \langle {\cal J}\left[A+{J\over {2\omega}}\right]
  \rangle. 
\nonumber \\
\end{eqnarray}
The second term originates from the square of the
magnetic field, $B^\sigma = [\nabla \times]^\sigma\mbox{}_\gamma A^\gamma +
V^\sigma\mbox{}_{\alpha\beta} A^\alpha A^\beta$, 
and it is shown in Fig.~5, 
\begin{eqnarray}
E_K/{\cal V} & = &  {1\over 2} \int [d\q] {{\omega^2(q) + \q^2}\over
 {2\Omega(q)}} \nonumber \\
& &  +  {g^2N_C\over 32}\int [d\q][d\k]  {{(3 - (\hat\k\cdot\hat\q)^2)}\over
 {\Omega(|\k|)
\Omega(|\q|) }}.
\end{eqnarray}
 \begin{figure}
 \includegraphics[width=2.5in]{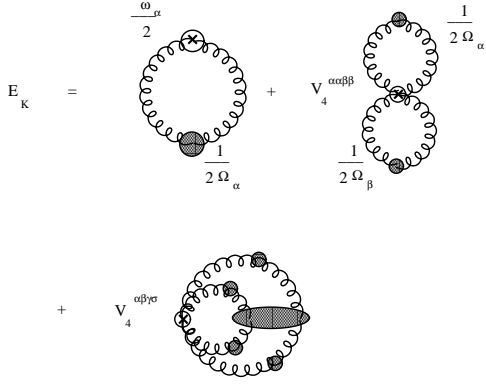}
 \caption{\label{fig5} The vacuum expectation value of the kinetic
   energy. The 4-point function contribution comes from the $\B^2$
   term}  
 \end{figure}
This  magnetic contribution involves a transverse gluon 4-point
function, which as discussed earlier, is approximated by the product 
 of two 2-point functions, {\it i.e.} the gluon-gluon scattering 
 amplitude shown in  Figs.~6  is not computed. 
% \begin{figure}
% \includegraphics[width=2.5in]{Fig4.eps}
% \caption{\label{fig1} Decomposition of the transverse gluon 4-point
%   function into a product of two 2-point functions and the 2-particle
%  (2PI) irreducible gluon scattering amplitude. }
% \end{figure}
 \begin{figure}
 \includegraphics[width=3.5in]{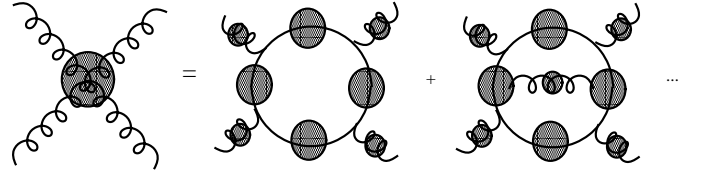}
 \caption{\label{fig6} Low order contributions to the 2PI gluon-gluon
   4-point function.}
 \end{figure}
The Coulomb energy {\it vev} is shown in Fig.~8 and is given by, 
\begin{eqnarray}
E_C & = &  {1\over 2}
  {\partial \over {\partial J_\alpha}} 
 {\partial \over {\partial J_\beta}} 
 {\partial \over {\partial J_\gamma}} 
 {\partial \over {\partial J_\delta}}\left[
 e^{ \sum_\rho {{ J^\rho J_\rho}\over
 {4\omega_\rho }}}
 \right. \nonumber \\
& & \left. \times   \rho_{\alpha\beta} \langle 
 (1 - \lambda)^{-2} (-\bm{\nabla}^2)
 {\cal J}\left[A+{J\over {2\omega}}\right] \rangle
  \rho_{\gamma\delta}   \right] \nonumber \\
\end{eqnarray}
with the charge density given by, Eq.~(\ref{rho}), $\rho = \rho^{\bar \gamma} = 
\rho^{\bar \gamma}\mbox{}_{\alpha\beta} A^\alpha A^\beta$,  and 
\begin{eqnarray}
E_C/{\cal V}  = {N_C\over 32} & & \int [d\k][d\q] 
{{K(|\q-\k|)}\over {(\q-\k)^2}}(1 + (\hat\k\cdot\hat\q)^2) \nonumber
\\
& & \times
{{ (   \Omega(|\k|)-F(|\k|) -\Omega(|\q|) + F(|\q|))^2}
\over {\Omega(|\k|)\Omega(|\q|)}}.  \nonumber \\
\end{eqnarray}
 \begin{figure}
 \includegraphics[width=2.5in]{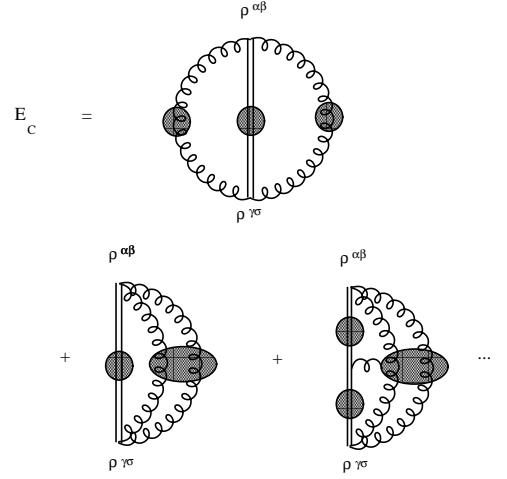}
 \caption{\label{fig7} The vacuum expectation value of the Coulomb operator}
 \end{figure}
Finally the gap equation follows from, 
$0 = \partial(E_K + E_C)/\partial\omega_\alpha$, 
\begin{widetext}
\begin{eqnarray}
\Omega^2(q) - F^2(q) - \q^2 & & = 
 {N_C g^2\over 4}\int [d\k] 
{{ (3 - (\hat\q\cdot\hat\k)^2)}
\over {\Omega(k)}}
\nonumber \\
& & + {N_C \over 4} \int [d\k] (1 + (\hat\q\cdot\hat\k)^2)
{{ K(\q-\k)}\over {(\q-\k)^2}} 
{{ (\Omega(k)-\Omega(q) - F(k) + F(q))
   (\Omega(q)+\Omega(k) - F(k) + F(q)) }
\over {\Omega(k)}}, \nonumber \\ \label{gap}
\end{eqnarray}
\end{widetext}
with 
\begin{equation}
K(q) = f(q) d^2(q),
\end{equation}
and $f$ satisfying 
\begin{equation}
f(q) = 1 + N_C \int [d\k] {{1 - (\hat\q\cdot\hat\k)^2}\over
  {2\Omega(|\k|)(\q-\k)^2}} f(|\q-\k|). \label{ff}
\end{equation}
It is also instructive to analyze the single quasiparticle dispersion
relation,
\begin{equation}
E_\alpha \delta^\alpha_\beta = \langle {\cal J } A^\alpha H A_\beta
\rangle/\langle {\cal J} \rangle.  
\end{equation}
The calculation is straightforward although more tedious due to
presence of up to three contractions corresponding the {\it vev} of
six field operators. The result is 
\begin{widetext}
\begin{eqnarray}
E(q) & =& {1\over {2\Omega(q)}}\left[ 
 \Omega^2 + F^2(q) + \q^2 
 +  {N_C g^2\over 4}\int [d\k] 
{{ (3 - (\hat\q\cdot\hat\k)^2)}
\over {\Omega(k)}} \right.
\nonumber \\
& & \left.+ {N_C \over 4} \int [d\k] (1 + (\hat\q\cdot\hat\k)^2)
{{ K(\q-\k)}\over {(\q-\k)^2}} 
{{ (\Omega(k)-\Omega(q) - F(k) + F(q))
   (\Omega(q)+\Omega(k) - F(k) + F(q))
  + 2\Omega^2(q) }
\over {\Omega(k)}} \right], \nonumber \\
\end{eqnarray}
\end{widetext}
After combining with the gap equation one obtains,
\begin{equation}
E(q) = \Omega(q)\left[ 1 +  
  {N_C \over 4} \int [d\k] (1 + (\hat\q\cdot\hat\k)^2)
{{ K(\q-\k)}\over {\Omega(k)(\q-\k)^2}} \right].
\end{equation}
which is identical to the expression found in Ref.~\cite{ss7} modulo 
 replacement  $\omega \to \Omega$.

\section{Renormalization}
So far we have been ignoring potential UV
 divergences. These divergences should be removed by renormalizing 
 appropriate operators and the coupling constant, $g$. It turns out
 that all four equations of interest,
 Eqs.~(\ref{dd}),~(\ref{Omega}),~(\ref{gap}),~(\ref{ff}),  
 require  renormalization. These equations have to be regularized
 first and this can be done by cutting-off the momentum integrals, 
 $\int [d\k] \to \int^\Lambda [d\k]$. The physical, renormalized
 solutions, $d(q)$,$f(q)$, $\omega(q)$ and $\Omega(q)$  should be $\Lambda$
 independent. We will first discuss renormalization of the expectation
 value of the FP operator, $d(q)$. Assuming that a renormalized
 solution for $\Omega(q)$ has been  found, the equation for $d(k)$ 
 is renormalized by adjusting the bare coupling, $g \to g(\Lambda)$, 
 {\it i.e.} the renormalized {\it vev} of the FP operator, $d(k)$ 
 will play the role of the running coupling. The $\Lambda$ dependence of
 $g(\Lambda)$ is determined by the UV behavior of Eq.~(\ref{dd}),  

\begin{equation}
{{dg(\Lambda)} \over {d\Lambda}} = -{\beta \over {(4\pi)^2}}
  {g^2(\Lambda) d(\Lambda)\over {\Omega(\Lambda)}}, \label{g}
\end{equation}
 with $\beta = 8N_C/3$.  In this and all other renormalization group
 equations  we keep only relevant and marginal
 contributions, {\it i.e.} no power corrections, $O(p^n/\Lambda^n)$
 with $n > 0$ are included since they do not require renormalization.  
 Since  $\beta > 0$ and physical solutions require $d(k),\Omega(k) > 0$ 
 the solution of Eq.~(\ref{g}) vanishes in the limit $\Lambda \to
\infty$.  In the limit $k\to \infty$ the integral 
 $I(k=\Lambda,\Lambda)$ given by Eq.~(\ref{dd}), with the second argument
 referring to the upper limit of integration,  is  finite.
 Thus in leading logarithmic approximation,  
  $d(\Lambda) \to g(\Lambda)$ as $\Lambda \to \infty$. 
   Furthermore, from Eqs.~(\ref{Omega}) it follows that  for large $q$, 
 $q\sim \Lambda$,  
\begin{equation}
{{d \Omega(q)} \over {dq}} \to {{d\omega(q)}\over {dq}}  +
O(d^2(q))  \to {{d\omega(q)}\over {dq}}  +
O(g^2). 
\end{equation}
Similarly from Eq.~(\ref{gap}), to leading logarithmic approximation, 
 we find $d\omega(q)/dq = 1 + O(g^2)$. Thus finally,  
\begin{equation}
\Lambda {{dg(\Lambda)} \over {d\Lambda}} = -{\beta \over {(4\pi)^2}}
 g^3(\Lambda) + O(g^5(\Lambda)), 
\end{equation}
which, ignoring the terms $O(g^5)$ has a solution given by 
\begin{equation}
g(\Lambda) = {{g(\mu)} \over {\left( 1 + {\beta\over {(4\pi)^2}} g^2(\mu) 
 \log(\Lambda^2/\mu^2)\right)^{1/2} }}. 
\end{equation}
The asymptotic behavior as $\Lambda \to \infty$ is therefore given by, 
\begin{equation}
g(\Lambda) =  {{4\pi} \over {\beta^{1/2}\log^{1/2}(\Lambda^2)}}.
\end{equation}
The renormalized equation for $d(k)$ is completely specified once
 $g(\mu)$, the value of the coupling at an arbitrarily chosen 
 renormalization scale, $\mu$ is fixed. It should be stressed, however,
  that this solution is valid only to within terms of the 
  order of $1/\log^{3/2}(\Lambda^2/\mu^2)$. 
 In practical applications we will be renormalizing at a low
  energy scale, $\mu$ {\it e.g.} related to the string tension or the
 glueball mass and thus such corrections become unimportant as
 $\Lambda\to \infty$.  For relevant operators, however, as we will see below, 
 such logarithmic corrections are multiplied by positive powers of 
 $\Lambda$ and thus cannot be neglected. 

 For practical (numerical) applications we have found a different,
  momentum subtraction renormalization (MSR) scheme to be more
  practical. In this scheme the renormalized equation for
  $d(q)$ is obtained by subtracting Eq.~(\ref{dd}) at $q=\mu$,   
\begin{widetext}
 \begin{equation}
 {1\over {d(q)}} - {1\over {d(\mu)}} 
 =   - {\beta\over
  {(4\pi)^2}} \int_{-1}^{1}  (\hat\k\cdot\hat\q)
 \int_0^\infty dk k^2 
{3\over 4}  {{(1 - (\hat\k\cdot\hat\q)^2)} \over {\Omega(k)}} 
 {{d(\q-\k)}\over {(\q-\k)^2}} 
 + (q \to \mu). \label{dr}
\end{equation}
\end{widetext}
In this renormalization scheme the coupling constant is therefore given by 
\begin{eqnarray}
& & {1\over {g_{MSR}(\Lambda)}} = {1\over {d(\mu)}}    \nonumber \\
& & 
 + {\beta\over {(4\pi)^2}}
 \int_{-1}^{1}
d(\hat\k\cdot\hat{\bm{\mu}})
 \int_0^\Lambda dk k^2 
{3\over 4} {{(1 - (\hat\k\cdot\hat{\bm{\mu}})^2)} \over {\Omega(k)}} 
 {{d(\bm{\mu}-\k)}\over {(\bm{\mu}-\k)^2}}, \nonumber \\
\end{eqnarray}
and $g(\Lambda) = g_{MSR}(\Lambda)$ to within corrections of the order
 $O(\mu/\Lambda)$ {\it i.e.} they agree asymptotically. From now on we
 will drop the $MSR$ subscript.  Finally for $k \sim \Lambda \to \infty$ 
\begin{equation}
d(k) = g(\Lambda)\left[1  + O\left( g^2(\Lambda) \log(\Lambda^2/k^2)
  \right) \right], \label{corrd}
\end{equation}
which is expected, as discussed above. 

We now proceed to discuss renormalization of the equation for
$f(k)$. Physically $f(k)$ represents additional contributions to the 
 {\it vev} of the square of the FP operator, $\langle (1-\lambda)^{-2} 
 \rangle \sim d^2 f$  not present in the expectation value, 
 $\langle (1-\lambda) \rangle^{-2} \sim d^2$. Any UV divergent
 contribution to $f$ should therefore be renormalized by renormalizing
 the operator $g^2/(1-\lambda)^2$, since the operator
 $g/(1-\lambda)$, has already being renormalized. This is done by
  a renormalization constant, $Z_K(\Lambda)$ introduced by replacing 
 the composite Coulomb operator, $K[A] \to Z_K(\Lambda) K[A]$. 
 Using the renormalized Coulomb operator the  equation Eq.~(\ref{ff}) for 
 $f(k)$ becomes, 
\begin{widetext}
\begin{equation}
f(k) = Z_K(\Lambda) + {\beta\over {(4\pi)^2}} 
 \int_{-1}^1  (\hat\k\cdot\hat\q)
\int_0^\Lambda dk k^2 
{3\over 4} {{(1 - (\hat\k\cdot\hat\q)^2)} \over {\Omega(k)}} 
 {{d^2(\q-\k)f(\q-\k)}\over {(\q-\k)^2}}, 
\end{equation}
\end{widetext}
and in the limit $\Lambda \to \infty$ one obtains, 
\begin{equation}
\Lambda {{dZ_K(\Lambda)}\over {d\Lambda}} = -{\beta\over {(4\pi)^2}} 
  {{d^2(\Lambda) f(\Lambda)}\over
 {\Omega(\Lambda)}} 
  Z_K(\Lambda),
\end{equation}
which in the leading logarithmic approximation has a solution given by 
\begin{equation}
Z_K(\Lambda) =  { {Z_K(\mu)}\over {\log^{1/2}(\Lambda^2/\mu^2)}}.
\end{equation}
 Choosing a value for $Z(\mu)$ at some UV point fixes the
 renormalized equation for $f(k)$. As in the case of the FP
 determinant, we will employ the momentum subtraction renormalization scheme,
 which leads to, 
\begin{widetext}
\begin{equation}
f(k) - f(\mu) =   {\beta\over {(4\pi)^2}} 
\int_{-1}^1 (\hat\k\cdot\hat\q) 
\int_0^\Lambda dk k^2 
{3\over 4} {{(1 - (\hat\k\cdot\hat\q)^2)} \over {\Omega(k)}} 
 {{d^2(\q-\k)f(\q-\k)}\over {(\q-\k)^2}}  - (q \to \mu), \label{fr}
\end{equation}
\end{widetext}
resulting, in the MSR scheme  in $Z_K(\Lambda)$ given by, 
\begin{widetext}
\begin{equation}
 Z_K(\Lambda) = f(\mu) -
 {\beta\over {(4\pi)^2}} 
 \int_{-1}^1 (\hat\k\cdot\hat{\bm{\mu}}) 
\int_0^\Lambda dk k^2 
 {3\over 4} {{(1 - (\hat\k\cdot\hat{\bm{\mu}})^2)} \over {\Omega(k)}} 
  {{d^2(\bm{\mu}-\k)f(\bm{\mu}-\k)}\over {(\bm{\mu}-\k)^2}}.  
\end{equation}
\end{widetext}
As expected, for UV values of $k\sim \Lambda \to \infty$ we find
\begin{equation}
f(k) = Z_K(\Lambda)\left[1  +
  O\left(g^2(\Lambda)\log(\Lambda^2/k^2)\right) \right].
\end{equation}
Much of the discussion on renormalization of $d$ and $f$ has already
being given in Ref.~\cite{ss7}. It should be noted that if for large
$k$, $d^2(k)/\Omega(k) < 1/\log(k)^n$ with $n>1$ there is no
renormalization for $f(k)$. Our analysis suggests that $n=1$ thus the
unrenormalized equation for $f$ has a sub-leading $\log(\log(k))$. We
suspect that this is an artifact of the rainbow-ladder truncation.  

As long as one works with the 
 leading logarithmic approximation, $\Omega(k) = \omega(k) = k$, 
 and  there is no effect of the FP determinant on $d$ or $f$. The 
 inclusion of the FP determinant has an effect on the low momentum behavior 
 of $d$ and $f$,  but it also introduces a new divergent integral, 
$F(q)$ in Eq.~(\ref{Omega1}). Since the origin of $F$ is the FP determinant 
${\cal J}$, it is the FP determinant that has to be renormalized in
 order to make $\Omega(k)$ finite. The 
 renormalized FP determinant should by chosen as,  
\begin{equation}
{\cal J} \to \left[{\cal J} e^{\sum_\alpha \delta \omega_\alpha A^\alpha
    A_\alpha + \cdots }\right]_\Lambda. \label{Jr} 
\end{equation}
 Here $\cdots$ stands for higher powers of the field operators,
 however,  since including the FP determinant within a
  gaussian approximation to the functional integrals only the
 quadratic term needs to be retained. It can be easily verified that 
 replacing ${\cal J}$ by Eq.~(\ref{Jr}) leads to the replacement, 
\begin{equation}
F(q) \to F(q,\Lambda) + \delta \omega (\Lambda),
\end{equation}
where $F(q,\Lambda)$ stands for the integral in Eq.~(\ref{Omega1}) with the
upper limit set to $\Lambda$. The counterterm $\delta \omega(\Lambda)$
 will be chosen to make $\Omega(q)$ UV finite. 
  Since $F(q)$ has mass dimension of one, in general one expects 
  two counter-terms would be needed, one proportional to $\Lambda$ and 
  the other to one power of the momentum. From the UV behavior of the
  integrand in Eq.~(\ref{Omega1}) it follows, however, that only the
  first is needed and we obtain, 
 \begin{equation}
{{d \delta\omega(\Lambda) }\over {d\Lambda}}  = - {\beta\over {(4\pi)^2}} 
 d^2(\Lambda),
\end{equation}
whose solution is given by 
\begin{equation}
\delta \omega(\Lambda) = \delta\omega(\mu) -{{\beta}\over {(4\pi)^2}} 
  \int_\mu^\Lambda d k d^2(k).
\end{equation}
 We note here that corrections to the leading asymptotic behavior 
  $d(k) \sim g(k)$ cannot be neglected here since 
 for $F(q)$ they result in terms of $O(\Lambda)$. Thus it is necessary
 to keep $d(k)$ rather then $g$ in the renormalized expression for
 $F$.   As in the case of $d$ and $f$ in the following we will use 
 the  MSR scheme for $\Omega(q)$ which gives, 
\begin{widetext}
 \begin{equation}
\Omega(q) - \Omega(\mu) - \omega(q) + \omega(\mu) 
 =   {\beta\over {(4\pi)^2}} \int_{-1}^1 
 (\hat\k\cdot\hat\q) 
 \int_0^\infty dk k^2 
{3\over 4} {{(1 - (\hat\k\cdot\hat\q)^2)} \over {(\q -\k)^2}} 
 d(\k) d(\q-\k) - (q \to \mu), \label{Or}
\end{equation}
\end{widetext}
with asymptotic behavior, $k\sim \Lambda \to \infty$  given by,
\begin{equation}
\Omega(k) = \omega(k) + {\beta \over
  {(4\pi)^2}} g^2(\Lambda)
\Lambda \left[ 1 +
  O\left(g^2(\Lambda)\log(\Lambda^2/k^2)\right)\right], 
\end{equation}
and with $\delta\omega(\Lambda)$ in MSR given by, 
\begin{eqnarray}
& & \delta \omega(\Lambda) = \Omega(\mu) - \omega(\mu) \nonumber \\
& & - 
 {\beta\over {(4\pi)^2}} \int_{-1}^1 
 (\hat\k\cdot\hat\q) 
 \int_0^\Lambda dk k^2 
{3\over 4} {{(1 - (\hat\k\cdot\hat{\bm{\mu}})^2)} \over {(\bm{\mu} -\k)^2}} 
 d(\k) d(\bm{\mu}-\k). \nonumber \\
\end{eqnarray}

The gap equation is the one which cannot be  renormalized in a simple
 way like the previous equations. This
 is due to inconsistencies in the approximation 
 used. Specifically, the gap equation is derived by taking the
 functional derivative of the energy expectation value with respect to
 $\omega$. In the second integral in Eq.~(\ref{gap}) we have
 retained only the derivative of the gluon lines and not of the Coulomb
 kernel. The former leads to terms in the integrand 
 proportional to $F$  thus formally of $O(g^4)$. 
 Similarly derivatives of the Coulomb operator $d^2f$  lead to terms 
 proportional to $d^4 f/\Omega^2$ {\it i.e.} also of
 $O(g^4)$. Thus if terms proportional to the difference $\Omega -
 \omega$ are kept in the numerator of the gap equation it would be
 necessary to include derivatives of the Coulomb kernel. However
 all these $O(g^4)$ terms involve two-loop integrals and for
 simplicity will be neglected. The simplified gap equation then reads, 
\begin{widetext}
\begin{equation}
\omega^2(q)   = \q^2 + 2 F(q) \omega(q) 
+ {N_C g^2\over 4}\int [d\k] 
{{ (3 - (\hat\q\cdot\hat\k)^2)}
\over {\omega(k)}}
 + {N_C \over 4} \int [d\k] (1 + (\hat\q\cdot\hat\k)^2)
{{ K(\q-\k)}\over {(\q-\k)^2}} 
{{ \omega^2(k)-\omega^2(q) } 
\over {\omega(k)}},  \label{gapnew}
\end{equation}
\end{widetext}
{\it i.e.} is the same as in Ref.~\cite{ss7} except for the term 
involving $F$ in the {\it r.h.s} and all terms are of $O(g^2)$. 
  This simplification is justifiable since our goal
 is to study the effect of the FP determinant on the low
momentum properties and thus possible modifications of 
 UV behavior are largely irrelevant.

In a covariant formulation the renormalized theory has the same
 operator structure as the bare one. This is not the case in the
 Hamiltonian approach. Renormalization introduces non-canonical 
 operators. The strength of such operators is determined by
  the cutoff. The gap equation has a quadratic divergence which is to be 
 renormalized by a gluon ''mass'' counter-term in the Hamiltonian,  
\begin{equation}
\delta H(\Lambda)  = {1\over 2} m^2(\Lambda)  A^\alpha A_\alpha.  
\end{equation}
 This is the only relevant operator {\it e.g.} of dimension two. 
  The constant $m^2(\Lambda)$ is fixed by requiring that the gap
 equation leads to a $\Lambda$-independent solution. Thus we insist 
  that $\omega(k)$ is $\Lambda$-independent and this guarantees that any
 divergence of an operator matrix element calculated with 
 respect to the state  $|\omega\rangle$ will be associated with the
 operator itself and not the sate. The counter-term $\delta H(\Lambda)$ 
  contributes to the {\it r.h.s} of Eq.~(\ref{gapnew}) with 
 $m^2(\Lambda)$ and from the UV behavior of Eq.~(\ref{gapnew}) we find,
\begin{equation}
{{dm^2(\Lambda)} \over {d\Lambda}} 
 = -{\beta\over{(4\pi)^2}} \left[ 2g^2(\Lambda) 
  {{\Lambda^2} \over {\omega(\Lambda)}}   
 + d^2(\Lambda) f(\Lambda) \omega(\Lambda)  \right],
 \end{equation}
whose solution is 
\begin{eqnarray}
& & m^2(\Lambda) - m^2(\mu) = \nonumber \\
& & = - {\beta\over{(4\pi)^2}} 
\int_\mu^\Lambda {{dk}\over {\omega(k)}}  \left[ 2g^2(\Lambda) 
  k^2 
  + d^2(k) f(k)\omega^2(k) 
  \right]. \nonumber \\
\end{eqnarray}
In the MSR scheme the gap equation then becomes, 
\begin{widetext}
\begin{eqnarray}
  \omega^2(q)  - 
  \omega^2(\mu) -  q^2  - F(q)\omega(q) &+& F(\mu)\omega(\mu)
 +   \mu^2  = \nonumber \\
& &  + {\beta\over {(4\pi)^2}}
\int_{-1}^1 (\hat\k\cdot\q)
 \int_0^\infty dk k^2 \left[\omega^2(k) - \omega^2(q)\right]
 {3\over 8}{ {(1 + (\hat\k\cdot\q)^2)}\over {\omega(\k)}} 
{{K(\q-\k)}\over {(\q-\k)^2}} 
 - (q \to \mu), \nonumber \\
\label{or}
 \end{eqnarray}
\end{widetext}
with the asymptotic behavior for $k\sim \Lambda \to \infty$ given by, 
\begin{eqnarray}
& & \omega^2(k) = q^2\left[ 1 + 
O\left(g^2(\Lambda)\log(\Lambda^2/k^2)\right)\right] \nonumber \\
& &  + {\beta\over {(4\pi)^2}} g^2(\Lambda)Z_K(\Lambda) \Lambda^2, 
\end{eqnarray}
and the counter-term $m^2(\Lambda)$ in the MSR is given by, 
\begin{widetext}
\begin{eqnarray}
m^2(\Lambda) &  = & 
 \omega^2(\mu) 
+  \mu^2   \nonumber \\
& &  - 2  {\beta \over {(4\pi)^2}}g^2(\Lambda) 
\int_0^\Lambda dk k^2{1\over {\omega(k)}}
  - {\beta\over {(4\pi)^2}} 
\int_{-1}^1 (\hat\k\cdot\hat{\bm{\mu}})
 \int_0^\Lambda dk k^2 \left[\omega^2(k) - \omega^2(\mu)\right]
 {3\over 8}{ {(1 + (\hat\k\cdot\hat{\bm{\mu}})^2)}\over {\omega(\k)}} 
{{K(\bm{\mu}-\k)}\over {(\bm{\mu}-\k)^2}}. \nonumber \\
\end{eqnarray} 
\end{widetext}
 The renormalized equations for the {\it vev} of the FP operator, the
  corrections to $d^2$ needed to obtain the Coulomb potential, the 
 gluon propagator and the ground state wave function are given by 
 Eqs.~(\ref{dr}),~(\ref{fr}),~(\ref{Or}), and ~(\ref{or}), respectively. 
 These equations depend on four parameters, the renormalization constants,
  $d(\mu)$, $f(\mu)$, $\Omega(\mu)$ and $\omega(\mu)$. In the
  following section we will study the solutions of these equations and
  their physical interpretation.

\section{ Results } 
As discussed above  there are four constants which need to be fixed. 
 This can be done, 
  for example, by comparing the Coulomb potential in position space, 
\begin{equation}
V_{eff}(\x) = \int {{d^3\k}\over {(2\pi)^3}} e^{i\k\cdot\x} 
 {{d^2(\k) f(\k)} \over {\k^2}} 
\end{equation}
 with the lattice, static quark-antiquark potential. This procedure
 was used in Ref.~\cite{ss7}. Unfortunately, the dependence of 
 $V_{eff}$ on the renormalization constants is complicated thus fitting
 the lattice potential will not necessarily provide much physical insight. 
 Furthermore it has recently being shown that there are differences 
 between the lattice Coulomb potential and the Wilson loop 
potential~\cite{Greensite:2003xf}. 
 We will thus proceed by simplifying the resulting equations and 
 imposing constraints on the renormalization constants. 
 The main difference between present analysis and what was done
 in Ref.~\cite{ss7} has to do with inclusion of the Faddeev-Popov
 determinant.  Our goal here is to investigate the role of the FP 
 determinant through the behavior of $\Omega(k)$ at low momenta. 
 If the determinant is omitted, one 
 has $\Omega(k) = \omega(k)$ and in this case the remaining three 
 equations, for $d(q)$, Eq.~(\ref{dr}),  $f(q)$, Eq.~(\ref{fr}) and $\omega(q)$,
 Eq.~(\ref{or}), were analyzed in Ref.~\cite{ss7}. 
 These equations have solutions provided $\omega(k)$ is finite as
 $k\to 0$. If $\omega(k) \to 0$ then the equation for $d(k)$ will
 develop a pole at a finite, positive value of momentum and if $\omega(k)
 \to \infty$ as $k\to 0$ then for a confining potential, $K(k) \to
 1/k^\alpha$, with $\alpha > 2$ the gap equation has no solution. 
 A renormalization condition at $\mu=0$, $\omega(0) = m_g$ was
 therefore imposed with $m_g$ fixed by the Wilson loop string
 tension. A simplified set of equations can be obtained by making an
 angular approximation,  
\begin{equation}
|\q - \k| \to q \theta(q - k) + k \theta(k-q)
\end{equation}
where $q=|\q|$ and $k=|\k|$. In Ref.~\cite{ss7} it was shown that the
angular approximation lead to results which are very close to the exact
numerical solutions. We will thus  follow this approximation here since it
 allows us to considerably simply the numerical analysis. 
 Using the angular approximation the equation for the FP operator, 
 $d(k)$ becomes, 
\begin{widetext}
\begin{equation}
{1\over {d(q)}} - {1\over {d(\mu)}} =
 -{\beta \over {(4\pi)^2}} 
 \int_0^q dk {{k^2}\over {q^2}} {{d(q)}\over {\Omega(k)}} 
+{\beta\over {(4\pi)^2}} \int_\mu^q dk {{d(k)}\over {\Omega(k)}} 
 + {\beta \over {(4\pi)^2}} \int_0^\mu dk {{k^2}\over {\mu^2}} 
{{d(\mu)}\over {\Omega(k)}}, \label{daa} 
\end{equation}
\end{widetext}
 the gluon propagator function, $\Omega$ is given by 
\begin{widetext}
\begin{equation}
\Omega(q) =\Omega(\mu) +  \omega(q)-\omega(\mu)  + {\beta \over {(4\pi)^2}} 
\int_0^q dk {{k^2}\over {q^2}} d(k)d(q) + 
{\beta \over {(4\pi)^2}}\int_q^\mu dk d^2(k)
- {\beta\over {(4\pi)^2}} \int_0^\mu dk {{k^2}\over {\mu^2}}
d(k)d(\mu), \label{Oaa}
 \end{equation}
\end{widetext}
and the  gap equation for $\omega(q)$ becomes, 
\begin{widetext}
\begin{eqnarray}
  \omega^2(q)  - 
  \omega^2(\mu) -  q^2  &-& F(q)\omega(q) + F(\mu)\omega(\mu)
 +   \mu^2  = \nonumber \\
& &  + {\beta\over {(4\pi)^2}}
 \int_0^q dk {{k^2}\over {q^2}} K(q) {{\omega^2(k) - \omega^2(q)}
 \over {\omega(k)}} 
 + 
 \int_q^\infty dk K(k) {{\omega^2(k) - \omega^2(q)}
 \over {\omega(k)}} 
 - (q \to \mu). \nonumber \\
\label{oaa}
 \end{eqnarray}
\end{widetext}
As discussed above if one ignores the FP determinant it is not possible
to choose an arbitrary renormalization condition for
$\omega(0)$. Furthermore in this case there is a critical (maximum)
  value of $d(\mu)=d_c(\mu)= 4\pi \sqrt{3\beta \omega(0)/5 \mu}$ for
 which Eq.~(\ref{daa}) has a solution. Appearance of such a critical
 coupling is an artifact of the approximation (rainbow-ladder
 truncation) used in evaluation of the expectation value of the
 Hamiltonian. The problem can be illustrated by considering the
 following integral, as a schematic representation of the 
 function integrals representing the ${\it vev}$ of the FP operator, 
\begin{equation}
I(g) = \int dx {\cal J}(x){ {e^{-\omega x^2}}\over {1 - gx}}. 
\end{equation}
Here $x$ represents the gauge potential, $J(x) \sim e^{log(1-x)}$ plays
  the role of the FP determinant, and $1/(1 - gx)$ the FP operator.  
 It one sets ${\cal J} =1$ the integral becomes divergent at $x=1$ unless
 $g=0$. The rainbow-latter approximation can be though off as a
 procedure for evaluating the integral be expanding $1/(1-gx)$ in a power
 series in $gx$ and then integrating term by term  keeping only a
  subset of contributions. In particular retaining the two-point
  correlations means the following approximation, 
\begin{equation}
\int dx x^{2n} e^{-\omega x^2} \to\left[ \int dx x^2 e^{-\omega
    x^2}\right]^n = {1\over \omega^{3/2}}. 
\end{equation}
This results in the following result for $I(g)$, 
\begin{equation}
I(g)/I(0) \sim {1\over {1 - g^2/\omega}}
\end{equation}
which has  a critical coupling $g=g_c = \sqrt{\omega}$. The FP
 determinant, however, makes the integral well defined for all values of
$g$ and thus no critical coupling is expected in this case. This is also
what happens if Eqs.~(\ref{Oaa}) is taken into account. 
 As long as $\Omega(0) > 0$ the $q=0$ value of $\omega(q)$ does not
 play a role in determining the position of the pole in the FP operator,
  and we can for simplicity assume $\omega(q) = 0$. Then
 from Eq.~(\ref{daa}) approximation  $\Omega(q) = \Omega(0)$  for $q <
 \mu$, we obtain, 
\begin{equation}
d(q) = {{ d(0) } \over {\left( 1 + {5\over 3}{\beta\over {(4\pi)^2}}
    d^2(0) q/\Omega(0) \right)^{1.2} } }. 
\end{equation}
Finally from Eq.~(\ref{Oaa}) we can derive a relation between $\Omega(0)$
and $d(\mu)$, 
\begin{equation}
z_0 \Omega(0) = \mu {\beta\over {(4\pi)^2}} d^2(0) 
\end{equation}
 \begin{figure}
 \includegraphics[width=2.5in]{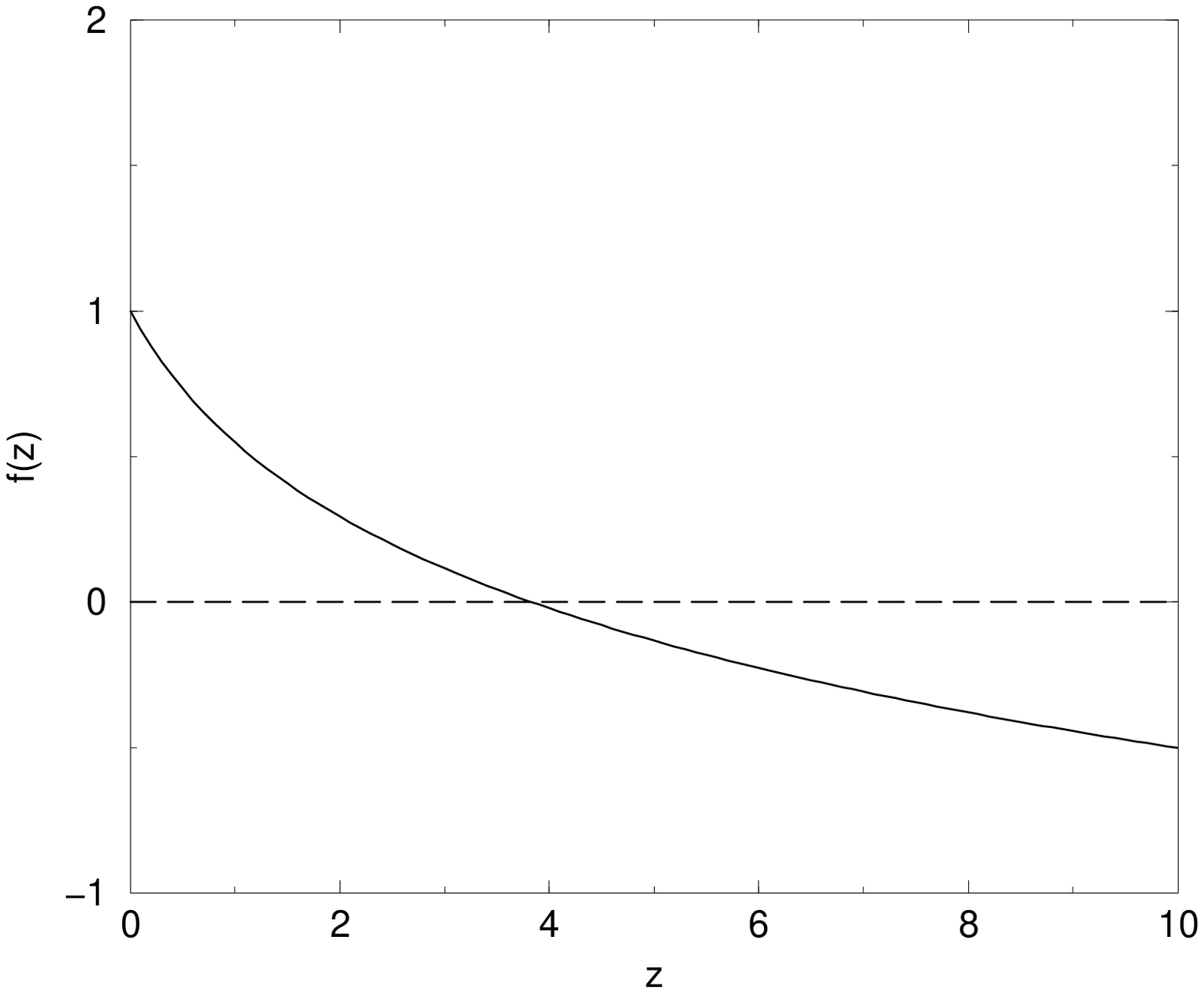}
 \caption{\label{fig8} Plot of the function $f(z) \equiv 1 - (\mbox{{\it
     r.h.s.} of Eq.~(\ref{Oaa})})/\Omega(0) = 
- 3\ln(1 + 5z/3)/5 + 31/25 - 25/125z + 144(1 - 1/\sqrt{1 +
  5z/3})/625z^2$, with $z \equiv d^2(\mu)\mu /\beta\Omega(0)$.
}
 \end{figure}
where $z_0\sim 4 $ is a  root of the nonlinear equation shown in Fig.~8
 Thus as $d(\mu)$ increases so does $\Omega(0)$ but there is no upper
 limit on $d(\mu)$. This is due to the FP determinant which
 regularizes functional integrals near the Gribov horizon. As 
 $\Omega(0)$ increases, the transverse-gluon 2-point correlation
 function decreases at low momentum and the ghost corelator function, 
 $d(k)$ increases. This is precisely what was found in other gauges
 using Dyson-Schwinger methods and in other approximations to the
 Coulomb gauge. 

 In Figs.~9-11 we plot results of numerical solutions to the set of
 coupled equations ~\ref{fr},~\ref{daa}, ~\ref{Oaa}, ~\ref{oaa}. 
 These should be compared with Figs.~4-6  from Ref.~\cite{ss7}.  In
 Fig.12 we plot the inverse of $\Omega(k)$ which is representing
 the transverse-gluon correlation function. As expected it is
 suppressed at low momenta and approaches the perturbative limit as $k\to
 \infty$.

 \begin{figure}
 \includegraphics[width=2.5in]{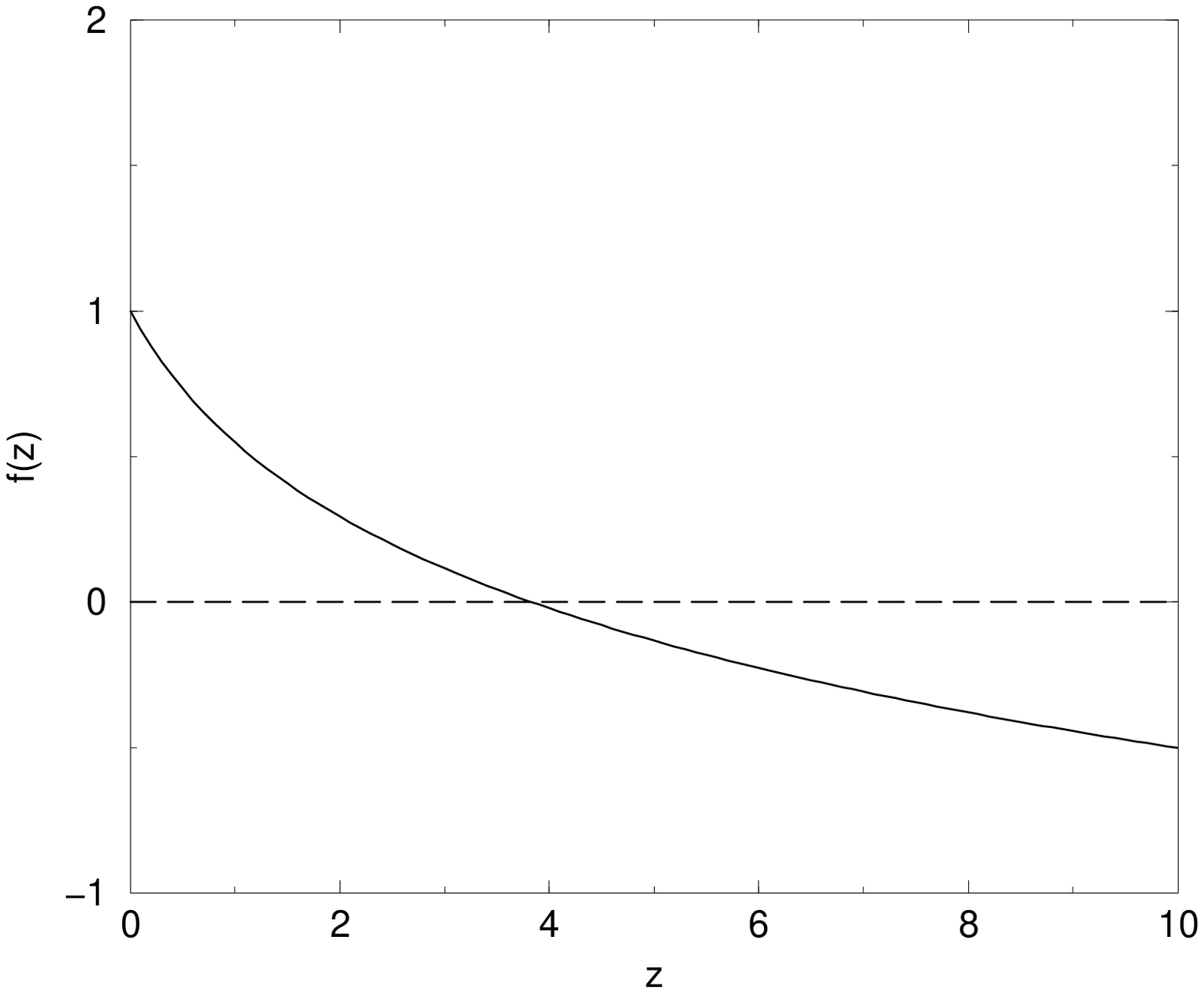}
 \caption{\label{fig9} A plot of the function $f(z) \equiv 1 - (\mbox{{\it
     r.h.s.} of Eq.~(\ref{Oaa})})/\Omega(0) = 
- 3\ln(1 + 5z/3)/5 + 31/25 - 25/125z + 144(1 - 1/\sqrt{1 +
  5z/3})/625z^2$, with $z \equiv d^2(\mu)\mu /\beta\Omega(0)$.
}
 \end{figure}

 \begin{figure}
 \includegraphics[width=2.5in]{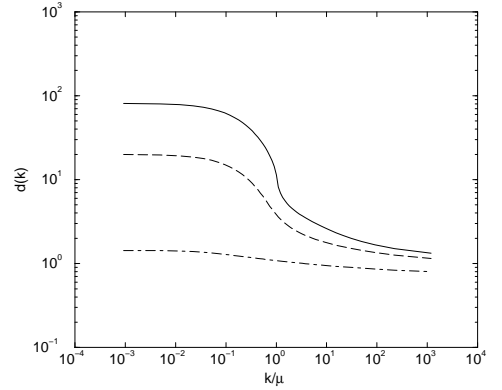}
 \caption{\label{d} Numerical solution for the vacuum expectation
   value of the FP operator, $d(k)$. The three curves correspond do
    $d(k=\mu)= 38.4$  (solid), $d(k=\mu) = 3.8$ (dashed) and $d(k=\mu)
    = 1.1$ (dashed-dotted) respectively, indicating that solution may
    exist for arbitrary choice of $d(\mu)$ {\it i.e.} no critical
    coupling. } 
 \end{figure}

 \begin{figure}
 \includegraphics[width=2.5in]{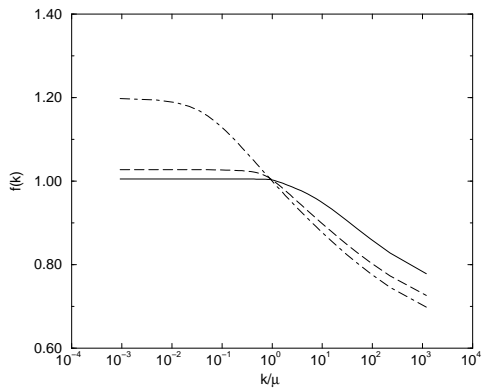}
 \caption{\label{fig9} Numerical solution for $f(k)$ normalized to 
   $f(k=\mu)= 1$. Labeling of curves is the same as in Fig.~\ref{d}.}
 \end{figure}

 \begin{figure}
 \includegraphics[width=2.5in]{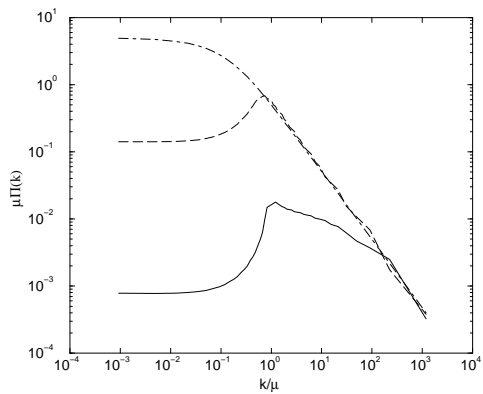}
 \caption{\label{fig9} Numerical solution for the instantaneous gluon
   propagator. Labeling of curves is the same as in Fig.~\ref{d}. 
 Increasing coupling $d(\mu)$ results in a stronger
   suppression at low moment.}  
 \end{figure}
   
\section{Summary}
In this paper we have studied the role of the Faddeev-Popov determinant in the
Coulomb gauge. The FP determinant specifies the measure in the
functional integrals over gauge field configurations and has so
 far been ignored in most calculations of QCD matrix elements in the Coulomb
gauge. The FP determinant vanishes at the boundary of the Gribov
region nevertheless it still allows for large field configurations near the
 boundary to enhance matrix elements. In particular we have shown that
the FP operator, corresponding to the running coupling and the ghost
 propagator is strongly enhanced in the IR, but at the same time no
artificial critical coupling exists. The same is true for the Coulomb
kernel which specifies the static, temporal Wilson loop. Finally,
the instantaneous part of the transverse gluon propagator is found to be
suppressed as is found in other gauges.

\section{Acknowledgment}
I would like to thank R.~Alkofer, H.~Reinhardt and D.~Zwanziger, for
 several discussions and S.~Teige for reading the manuscript.
  This work was supported in part by the US
Department of Energy grant under contract 
 DE-FG0287ER40365.

\end{document}